\documentstyle[multicol,prl,aps,epsf]{revtex}
\begin{document}
%\draft
\tightenlines
\title{Quantum Fluctuations and Resistivity of Thin Superconducting Wires}
\author{Andrei D. Zaikin$^{1,2}$, Dmitrii S. Golubev$^{2}$, 
        Anne van Otterlo$^{3}$, and Gergely T. Zimanyi$^4$}
\address{$^1$ Institut f\"{u}r Theoretische Festk\"orperphysik,
	Universit\"at Karlsruhe, 76128 Karlsruhe, FRG\\
        $^2$ I.E.Tamm Department of Theoretical Physics, P.N.Lebedev 
	Physics Institute, Leninskii pr. 53, 117924 Moscow, Russia\\
	$^{3}$ Theoretische Physik, ETH-H\"{o}nggerberg,
	CH-8093 Z\"{u}rich, Switzerland\\
        $^4$ Physics Department, University of California, Davis,
	CA 95616, USA}
\maketitle

\begin{abstract}
We present a microscopic study of the quantum fluctuations of the
superconducting order parameter in thin homogeneous superconducting wires
at all temperatures below $T_C$. The rate of quantum phase slip processes
determines the resistance $R(T)$ of the wire, which is observable in very thin
wires, even at low temperature. Furthermore, we predict a new low temperature
metallic phase below a critical wire-thickness in the 10nm range, in which
quantum phase slips proliferate.
\end{abstract}
\begin{multicols}{2}

The role of fluctuations for the superconducting transition in reduced
dimension is well known. Above the critical temperature $T_{C}$,
such fluctuations yield an enhanced conductivity~\cite{almt}.
Below $T_C$ thermally activated phase slips (TAPS) result in
a finite resistance of one-dimensional (1D) superconductors~\cite{lamh}.
Close to $T_{C}$ the experimental results~\cite{Exp} fully confirm the
theoretical predictions~\cite{lamh}. However, as the temperature is lowered
the number of TAPS decreases exponentially and no measurable resistance 
is predicted by the theory~\cite{lamh} at $T$ not very close to $T_C$.
Nevertheless, recent experiments by Giordano~\cite{Gio} clearly
demonstrate a notable resistivity of thin superconducting wires 
far below $T_{C}$. More recently, similar experimental results on (quasi-)1D
systems have been also reported by others~\cite{Her,Lin}. 

As these experimental findings cannot be understood within the TAPS
theory~\cite{lamh}, it is tempting to attribute them to quantum fluctuations
which generate quantum phase slips (QPS) in a 1D superconducting wire. A first
estimate~\cite{Gio} for the QPS rate $\propto\exp(-S_{QPS})$, however, leads
to the disappointing conclusion that $S_{QPS}$ roughly equals the number of
transverse channels $N_{Ch}=k^{2}_{F}S$ in the wire ($S=\pi r^{2}_{0}$ is the
cross section of the wire), which is very large even for the thinnest wires
used in the experiments~\cite{Gio} (e.g. for $r_{0}\sim 10^{-6}$ cm we have
$S_{QPS}\sim 10^2 \div 10^{3}$), and therefore QPS effects should be strongly
suppressed. This estimate is obtained from the formula $S_{QPS}\sim U_{QPS}/
\omega_{a}$, with energy barrier $U_{QPS}$ and attempt frequency $\omega_{a}
\sim\Delta$. Assuming $U_{QPS}$ to be the condensation energy $N(0)\Delta^{2}
/2$ in a volume $\xi_0 S$ during a time $\Delta^{-1}$, one obtains $S_{QPS}
\sim\xi_0 S N_{0}\Delta^{2}/2\Delta\sim k^{2}_{F}S/4\pi^2 \sim N_{Ch}$. A
similar estimate has been obtained using a phenomenological time dependent
Ginzburg-Landau (TDGL) free energy with second order time
derivatives~\cite{sm,Duan}. Furthermore, recently Duan~\cite{Duan} argued that
the electromagnetic field yields an additional suppression of the QPS rate
by the factor $\exp(-1/\alpha )$, $\alpha =1/137$ is the fine
structure constant. Even further suppression of this rate $-$ similar
to the case of Josephson junctions~\cite{cl} $-$ can be expected due to
dissipative currents in the QPS core. In contrast, the magnitude of the
resistance for the thinnest wires measured in Ref.~\cite{Gio} yields $S_{QPS}
\sim 10$ with the QPS rate by orders of magnitude larger than
that derived from the above estimates.

In this Letter we argue that the above estimate needs {\it qualitative}
improvement. First, the estimate for the potential barrier should be
corrected: as the typical electron mean free path in the wires~\cite{Gio} is
very small $l\lesssim 10$ nm, one should rather take $\xi\sim\sqrt{l\xi_{0}}$ as
the typical QPS size. This decreases the estimated value of $S_{QPS}$ by one
order of magnitude. Second, we show below that the role of the electromagnetic
field for thin  wires was overestimated in Ref.~\cite{Duan} (roughly by a
factor $r_{0}/\lambda_{L} \sim 10^{-1}\div 10^{-2}$, $\lambda_L$ is the London
length of a bulk superconductor). Third, the dissipative currents turn out not
to have a strong impact on the QPS rate, especially in the limit of low $T$.
Also from a general point of view, TDGL-based theories~\cite{sm,Duan} are
insufficient at $T$ not very close to $T_C$ and fail to give qualitatively
correct results. A microscopic theory of QPS is needed that accounts for
non-equilibrium, dissipative and electromagnetic effects during a QPS event.

This theory is reported upon below. Taking into account the above effects we
$determine$ both the characteristic size $x_{0}$ and time scale $\tau_{0}$ of
a QPS core (in general these depend on dissipation and electromagnetic
characteristics of the wire and {\it do not} coincide with $\xi$ and
$1/\Delta$) and derive a microscopic expression for the QPS rate. At $T=0$ we
predict a new {\it Metal-Superconductor} ($MS$) phase transition governed by
the strength of electromagnetic interactions between different QPS. We
evaluate the effective resistance of a 1D superconducting wire, demonstrate
that in sufficiently thin wires QPS's are observable at all $T$, and determine
the crossover temperature between the regimes governed by TAPS and QPS.

{\it The model.}
Our calculation is based on the effective action approach for a BCS
superconductor~\cite{sz}. The starting point is the partition function $Z$ 
expressed as an imaginary time path-integral over the electronic fields $\psi$
and the gauge fields $V, {\bf A}$, with Euclidean action
\begin{eqnarray}\nonumber
	S= \int d^{3}{\bf r}\int^{\beta}_{0}d\tau {\Big (}
	\bar{\psi}_{\sigma}
	[\partial_{\tau}-ieV+\xi({\bf \nabla}-ie{\bf A}/c)]
	\psi_{\sigma}  - \\ \nonumber
	-g\bar{\psi}_{\uparrow}\bar{\psi}_{\downarrow}
	\psi_{\downarrow}\psi_{\uparrow} + ieVn_{i}+
	[{\bf E}^{2}+{\bf B}^{2}]/8\pi	{\Big )} \; .
\end{eqnarray}
Here $\beta =1/T$, $\xi({\bf \nabla})\equiv -{\bf \nabla}^{2}/2m - \mu$
describes a single conduction band, $g$ is the BCS coupling constant, $en_{i}$
denotes the background charge density of the ions, and units in which
$\hbar=k_{B}=1$ are used. A Hubbard-Stratonovich transformation introduces the
energy gap $\Delta$ as an order parameter and the electronic degrees of
freedom can be integrated out. What remains is an expression for the partition
function in terms of an effective action for $\Delta$, $V$ and ${\bf A}$, with
a saddle-point solution $\Delta=\Delta_{BCS}$ and $V={\bf A}=0$
\begin{eqnarray}\nonumber
	S_{\rm eff}=\int d^{3}{\bf r}\int^{\beta}_{0}d\tau
	\left[\frac{|\Delta|^2}{g}+\frac{{\bf E}^2+{\bf B}^2}{8\pi} \right]-
		\mbox{Tr}\ln\hat{G}^{-1} \; ,\\ \nonumber
	\hat{G}^{-1}=\left(\partial_{\tau}+\frac{i}{2}
	\{{\bf \nabla},{\bf v}_{s}\}\right)\hat{1}+\Delta_{0}\hat{\sigma}_{1}+
	\\ \nonumber
	+\left(\xi({\bf \nabla})+\frac{m{\bf v}^{2}_{s}}{2}-ie\Phi\right)
	\hat{\sigma}_{3}\;,
\end{eqnarray}
where the superfluid velocity ${\bf v}_{s}=(1/2m)[{\bf \nabla}\varphi-
2e{\bf A}/c]$, the chemical potential for Cooper pairs $\Phi=V-
\dot{\varphi}/2e$, and $\Delta=\Delta_{0}e^{i\varphi}$ have been introduced.

{\it Effective action.}
The effective theory is constructed by expanding up to second order around the
saddle point in $\Phi$ and ${\bf v}_{s}$ to obtain the electronic
polarization terms. Using the Ward identity from gauge invariance, the result
can be written as the sum of terms, related to the normal and superfluid
densities $n_{n}$ and $n_{s}$, that describe normal and superconducting
``screening'' respectively~\cite{ogzb,gozz}
$$
	S_{\rm pol}=\frac{S}{\beta}\sum_{\vert\omega_{\mu}\vert>\Delta_{0}}
	\int dx	\frac{\sigma}{2|\omega_{\mu}|}E^{2}+
$$ 
\begin{eqnarray}\nonumber
	+S\int dxd\tau\left(\frac{mn_{s}}{2}v^{2}_{s}+e^{2}N_{0}
	[\frac{n_{n}}{n}V^{2}+\frac{n_{s}}{n}\Phi^{2}]\right) \;,
\end{eqnarray}
where use was made of the one-dimensional nature $r_{0}<\xi$ of the problem,
and $\sigma$ is the conductivity of the wire.
Transverse screening is irrelevant if the London penetration depth
$\lambda_{L}>r_{0}$ and we retain only one component of the vector
potential~\cite{lamh}.

A phase-slip event in imaginary time involves a suppression of the order
parameter in the phase slip core and a winding of the superconducting phase
around this core~\cite{foot}. We now separate the total QPS action $S_{QPS}$
into a core part $S_{\rm core}$ around the phase slip center for which the
condensation energy and dissipation by normal currents are important (scales
$x\leq x_{0}$, $\tau\leq \tau_{0} $), and a hydrodynamic part outside the core
$S_{\rm out}$ which depends on the hydrodynamics of the electromagnetic fields
\begin{eqnarray}\nonumber
	S_{\rm out}=\!\int\! dxd\tau{\Big (} \frac{C+C'}{2}V^{2}+
	\frac{\tilde{C}}{2}\Phi^{2}+ \frac{1}{2Lc^{2}}A^{2}+
	\frac{m^{2}{\bf v}^{2}_{s}}{2e^{2}\tilde{L}} {\Big )} \; ,
\end{eqnarray}
where the kinetic inductance $\tilde{L}=m/(e^{2}n_{s}S)$ and kinetic
capacitance $\tilde{C}=Se^{2}N_{0}n_{s}/n$ have been introduced, as well as
$C'=Se^{2}N_{0}n_{n}/n$ which we will drop from now on in the limit
$n_{s}\gg n_{n}$ at low $T$. The geometry and screening by dielectrics outside
the wire are accounted for by the capacitance per length $C=\epsilon_{r}
[2\ln(x_{0}/r_{0})]^{-1}$ and the inductance times length $L=2\ln(x_{0}/r_{0})
/c^{2}$ that replace the ${\bf E}^{2}+{\bf B}^{2}$-term. Here $c$ is the
velocity of light and $\epsilon_{r}$ the dielectric constant of the substrate.

{\it Single QPS.}
Outside the phase slip core, the equations of motion for $V$, $A$, and
$\varphi$ are solved by the saddle point
\begin{eqnarray}\nonumber
	\tilde{\varphi}=\arg(x+ic_{0}\tau) \; & ; & \;\;
	c^{2}_{0}=\frac{C^{-1}+\tilde{C}^{-1}}{L+\tilde{L}}\\ \nonumber
	V=\frac{1}{1+C/\tilde{C}}\frac{\partial_{\tau}\varphi}{2e} \; & ; & \;\;
	A=\frac{c}{1+\tilde{L}/L}\frac{\partial_{x}\varphi}{2e} \; .
\end{eqnarray}
The space$-$time anisotropy is determined by the plasmon velocity $c_{0}$,
rather than by $v_{F}$. For generic parameters the velocity $c_{0}$ reduces to
the velocity of the Mooij-Sch\"{o}n mode, which has dispersion $\omega^{2}=
c^{2}_{MS}k^{2}$ with $c^{2}_{MS}= S\omega^{2}_{pl}/4\pi C$~\cite{ms}, where
$\omega_{pl}$ is the 3D plasma frequency. The corresponding saddle point
action is
\begin{eqnarray}
	S^{*}_{\rm out}=\mu\ln[\min(c_{0}\beta,X)/\max(c_{0}\tau_{0},x_{0})]\;,
\label{hydr}
\end{eqnarray}
with $\mu=\pi/[4e^{2}c_{0}(L+\tilde{L})]$.

The contribution from the core part is estimated to be
\begin{eqnarray}
	S^{*}_{\rm core}=\frac{N_{0}}{2}S\tau_{0}x_{0}\Delta^{2}_{0}+
	\frac{S}{\beta}\sum_{|\omega_{\mu}|>\tau^{-1}_{0}}\frac{x_{0}\sigma}
	{|\omega_{\mu}|}|E(\omega_{\mu},\frac{x_{0}}{2})|^{2} \; .
\label{Score}
\end{eqnarray}
The first part is the condensation energy that is lost inside the core and the
second part defines the energy of dissipative currents inside the core
during a phase slip event. Inserting the saddle point solution, and minimizing
the action with respect to $x_0$ and $\tau_{0}$, we find $x_{0}\approx
c_{0}\tau_{0}\approx(\sigma c_{0}/2e^{2}N_{0}\Delta^{2}_{0})^{1/3}$ and
$S^{*}_{core}$ as three times the condensation energy in~(\ref{Score}).
In the Drude limit $\sigma= 2e^2N_0v_Fl/3$, we obtain $x_{0}\approx
(\xi^2c_0/\Delta_0)^{1/3}$ 
and
\begin{eqnarray}
	S^{*}_{\rm core}=bN_{0}S(\Delta_{0}\xi)^{4/3}/c_{0}^{1/3},\;\;\;b\sim 1.
\label{Score2}
\end{eqnarray}

{\it QPS Interactions.}
The next step is to consider a gas of QPS's in a superconducting wire. We also
assume that an applied current $I$ (much smaller than the depairing current)
is flowing through the wire. Substituting the saddle point solution $\varphi=
\sum^{n}_{i}\tilde{\varphi}(x-x_{i},\tau-\tau_{i})$ into the action and
keeping track of the additional term $\int d\tau\int dx(I/2e)\partial_{x}
\varphi$~\cite{sz}, we find
\begin{equation}
	S_{n}=nS^{*}_{\rm core}-\mu\sum\limits_{i\neq j}\nu_{i}\nu_{j}
	\ln \biggl{(}\frac{\rho_{ij}}{x_{0}}\biggr{)}+
	\frac{\Phi_{0}}{c}I\sum\limits_i\nu_{i}\tau_{i}\;.
\label{mul}
\end{equation}
Here $\rho_{ij}=(c^{2}_{0}(\tau_{i}-\tau_{j})^2+(x_{i}-x_{j})^2)^{1/2}$
defines the distance between the i-th and j-th QPS in the $(x,\tau)$ plane,
$\nu_{i}=+1$ ($-1$) are the QPS (anti-QPS) ``charges'', and $\Phi_{0}=hc/2e$
is the flux quantum. Only neutral QPS configurations with $\nu_{tot}=
\sum_{i}^{n} \nu_{i}=0$ (and hence $n$ even) contribute to the partition
function. This is a consequence of the boundary condition
$\varphi(x,\tau)=\varphi(x,\tau+\beta)$ in the path integral for the partition
function~\cite{sz}.

{\it Metal-Superconductor phase transition.}
For $I=0$ Eq.~(\ref{mul}) defines the standard model of a 2D 
gas of logarithmically interacting charges $\nu_{i}$. The effective (small)
fugacity $y$ of these charges (or the QPS rate $y/x_0\tau_0$) is
\begin{equation}
	y=x_{0}\tau_{0}B\exp(-S^{*}_{\rm core})\;,
\label{fug}
\end{equation}
where $B$ is the usual fluctuation determinant with zero modes excluded. From
the Coulomb gas analogy, we conclude that a KTB phase transition~\cite{ktb}
for QPS's occurs in a superconducting wire at $\mu=\mu^{*}\equiv 2+4\pi y
\approx 2$: for $\mu<\mu^{*}$ the density of free QPS in the wire (and
therefore its resistance) always remains finite, whereas for $\mu >\mu^{*}$
QPS's and anti-QPS's (AQPS) are bound in pairs and the $linear$ resistance of
a superconducting wire is strongly suppressed and $T$-dependent. We
arrive at an {\it important conclusion}: at $T=0$ a 1D superconducting wire
$is$ in a superconducting state, with vanishing linear resistance, provided
the electromagnetic interaction between phase slips is sufficiently strong,
i.e. $\mu>\mu^{*}$.

The above analysis is valid for sufficiently long wires. For typical
experimental parameters, however, $X<c_{0}\beta$ (or even $X\ll c_{0}\beta$),
and the finite wire size needs to be accounted for. To this end, we first
apply the standard 2D scaling analysis~\cite{ktb} $\partial_{l}\mu=-4\pi^{2}
\mu^{2}y^2$ and $\partial_{l}y=(2-\mu)y$, where $\mu$ and $y$ depend on the
scaling parameter $l$. Solving these equations up to $l=l_{X}=\ln(X/x_{0})$ 
we obtain the renormalized values $\tilde{\mu}=\mu(l_{X})$ and $\tilde{y}=
y(l_{X})$. For larger scales $l>l_{X}$ only the time coordinate matters and
the problem reduces to a that of a 1D Coulomb gas with logarithmic
interaction. Therefore, (for $\tilde{y}\ll 1$) further scaling for $l>l_{X}$
is defined by \cite{s,sz} $\partial_{l}\mu=0$ and $\partial_{l}y=(1-\mu)y$.
For $\tilde{\mu}>1$ the fugacity scales down to zero, which again corresponds
to a superconducting phase, whereas for $\tilde{\mu}<1$ it increases
indicating a resistive phase in complete analogy to a single Josephson
junction with ohmic dissipation. Thus, our above conclusion about the presence
of a $MS$ phase transition at $T=0$ remains valid also for relatively short
wires, although the phase transition point is given by the somewhat different
condition $\tilde{\mu}=1$. In practice, both conditions $\mu>\mu^{*}$ and
$\tilde{\mu}=1$ are realized in wires with diameter $2r_{0}\sim 10\div 20$ nm,
see also the discussion below.

{\it Wire resistance at low T.}
At any nonzero $T$ the wire has a nonzero resistance $R(T,I)$ even in the 
``ordered'' phase $\mu >\mu^{*}$ (or $\tilde \mu >1$). In order to evaluate
$R(T)$ in this phase for a long wire we proceed perturbatively and first
calculate the free energy correction $\delta F$ due to one bound QPS-AQPS
pair. [See Ref.~\cite{sz} (Chapter 5.3) for a discussion of a similar
procedure.] The one QPS-AQPS pair contribution $\delta F$ to the free energy
of the wire is
\begin{equation}
	\delta F=\frac{Xy^{2}}{x_{0}\tau_{0}}\int^{\beta}_{\tau_{0}}
	\frac{d\tau}{\tau_{0}} \int^{X}_{x_{0}}\frac{dx}{x_{0}}
	e^{(\Phi_{0}I\tau/c)-2\mu\ln[\rho(\tau,x)/x_{0}]}\;,
\label{fren}
\end{equation}
where $\rho=(c^{2}_{0}\tau^{2}+x^{2})^{1/2}$. It is convenient to first
integrate over the spatial coordinate $x$ and take the wire length
$X\rightarrow\infty$. For nonzero $I$ the expression in Eq.~(\ref{fren})
is formally divergent for $\beta\rightarrow\infty$ and acquires an imaginary
part Im $\delta F$ after analytic continuation of the integral over the
temporal coordinate $\tau$~\cite{gw,sz}. This indicates a QPS-induced
instability of the superconducting state of the wire: the state with a zero
phase difference $\delta\varphi(X)=\varphi(X)-\varphi(0)=0$ decays into a
lower energy state with $\delta\varphi(X)=2\pi$. The corresponding decay rate
is $\Gamma_{2\pi}=2\mbox{Im}\delta F$. The rate for the opposite transition
(which is nonzero at nonzero $T$) is defined analogously with $I\to -I$. The
average voltage drop $V=(\Phi_{0}/c) [\Gamma_{2\pi}(I)-\Gamma_{2\pi}(-I)]$
across the wire is
$$
	V=\frac{\Phi_{0}Xy^{2}}{c\tau_{0}x_{0}}
	\frac{\sqrt{\pi}\Gamma(\mu-\frac{1}{2})}
		{\Gamma(\mu)\Gamma(2\mu-1)}
	\sinh\left(\frac{\Phi_{0}I}{2cT}\right)
$$
\begin{equation}
	\mid\Gamma\left(\mu-\frac{1}{2}+\frac{i}{\pi}
		\frac{\Phi_{0}I}{2cT}\right)\mid^{2}
	\left[\frac{2\pi\tau_{0}}{\beta}\right]^{2\mu-2}\;,
\label{volt}
\end{equation}
$\Gamma (x)$ is the Euler gamma-function. For the wire resistance 
$R(T,I)=V/I$ this yields $R\propto T^{2\mu-3}$ and
$R\propto I^{2\mu-3}$ for $T \gg \Phi_{0}I$ and $T\ll\Phi_{0}I$ respectively.
For thick wires with $\mu>\mu^{*}$, we expect a strong temperature dependence
of the resistivity. For thinner wires the temperature dependence of the
resistivity becomes linear at the transition to the disordered phase in which
our analysis is not valid. At $T\ll\Phi_{0}I/c$ we expect a strongly nonlinear
I-V characteristic $V\sim I^{a}$ in thick wires, and a universal
$a(\mu^{*})=2$ in thin wires at the transition into the resistive state with
$V\sim I$, i.e. $a=1$. Note that in contrast to the KTB transition in 2D
superconducting films, the jump is from $a=2$ to 1, instead of $a=3$ to 1.

For a short wire $X<c_{0}/T$ we again proceed in two steps. A 2D scaling
analysis yields the ``global'' parameters $\tilde{y}$, $\tilde{\mu}$, and the
microscopic cut-off $\tilde{\tau}_{0}=\tau_{0}X/x_{0}$. In analogy with the
resistively shunted Josephson junction~\cite{sz}, the voltage drop from the
imaginary part of the free energy reads
$$
	V=\frac{2\Phi_{0}\tilde{y}^{2}}{\Gamma(2\tilde{\mu})c\tilde{\tau}_{0}}
	\sinh\left(\frac{\Phi_{0}I}{2cT}\right)
	\vert\Gamma(\tilde{\mu}+\frac{i\Phi_{0}I}{2\pi cT})\vert^{2}
	\left[\frac{2\pi\tilde{\tau}_{0}}
		{\beta}\right]^{2\tilde{\mu}-1}\;,
$$
giving $R\propto T^{2\tilde \mu-2}$ and $R\propto I^{2\tilde \mu-2}$
respectively at high and low $T$. This result is valid for $\tilde{\mu}>1$ and
also for smaller $\tilde{\mu}$ at not very small $T$~\cite{sz}. At $T\to 0$ in
the metallic phase the resistance flattens off and becomes~\cite{sz}
\begin{equation}
	R=R_{q}/\tilde{\mu}\;, 
\label{Rshort0}
\end{equation}
where $R_{q}=\pi/2e^2\simeq 6.5$ k$\Omega$ is the quantum resistance. Note
that in this limit the size dependence of the resistance enters only through
the renormalized value $\tilde{\mu}$.

{\it Crossover temperature.}
The cross-over between the TAPS and the QPS regime occurs at a temperature
$T^{*}$ that can be estimated by comparing the corresponding resistivities
$R_{TAPS}$ and $R_{QPS}$. The standard result of Ref.~\cite{lamh} is
$R_{TAPS}=\beta\Omega(h/4e^{2}) \exp(-\beta\Delta F)$, with $\Delta
F=\sqrt{2}H^{2}_{c}S\xi/3\pi$, attempt frequency $\Omega=(X/\xi)
\sqrt{\beta\Delta F}\tau^{-1}_{s}$, and relaxation time $\tau^{-1}_{s}=
8(T_{C}-T)/\pi$. For thin wires with $\mu<S^{*}_{\rm core}$ the value $T^{*}$
which follows from a comparison of the exponents $\beta\Delta F$ and
$2S^{*}_{\rm core}$ is
\begin{equation}
	T^{*}=\frac{\Delta F}{2S^{*}_{core}}
	\approx \Delta_0^{2/3}c_0^{1/3}/\xi^{1/3}.
\label{crossover}
\end{equation}
The pre-exponential factor $B$ in Eq.~(\ref{fug}) can be estimated by matching
the pre-exponential factors of $R_{QPS}$ and $R_{TAPS}$ at $T=T^{*}$. We
postpone a more detailed analysis of the pre-exponent to a forthcoming
publication~\cite{gozz}.

{\it Discussion}.
For typical system parameters $k^{-1}_{F} \sim 0.2$ nm $<l\sim 7$ nm $<\xi
\sim 10$ nm $<\xi_{0}\sim \lambda_{L}\sim 100$ nm, we find that $L$ and
$\tilde{C}$ drop out of the problem and $\tilde{L}$ and $C$ determine the
physics (unless $T\sim T_{c}$ or $\epsilon_{r}\gg 1$, see Ref.~\cite{gozz}).
We will also take the length $X$ of the wire to be smaller than the localization
length, so that localization effects do not play a role.
Taking $r_{0}\sim$ 10 nm and $\epsilon_{r}$ = 1, we obtain the velocity
$c_{0}/c=c_{MS}/c\approx (r_{0}/6\lambda_{L})$,
$$
	\mu=(\pi\sqrt{\epsilon_{r}}/8\alpha)
	(r_{0}/\lambda_{L})\approx 50(r_{0}/\lambda_{L})\;,
$$ 
and $2S_{core}^{*}\lesssim 10$. This estimate demonstrates that $-$ in
contrast to previous studies~\cite{Duan} $-$ quantum fluctuations in thin
superconducting wires are not negligibly small and can be well observed 
in experiment. Furthermore, our estimate for the classical-to-quantum
crossover temperature Eq.~(\ref{crossover}) yields $T^{*}\sim 10\Delta(T^{*})$,
i.e. for thin wires one expects this crossover to happen quite close to $T_C$.
These features are in good agreement with the experimental
findings~\cite{Gio}.

For the quoted parameters, we predict the superconductor to metal transition
at a wire thickness of order $r_{0}=\lambda_{L}/25\approx 5\div 10$ nm. This
prediction agrees with the results of Giordano, who finds that wires with
$r_0 \approx 8$ nm have a resistivity that saturates at a measurable level 
at low $T$, whereas the resistivity of thicker wires $r_0 \gtrsim 13$ nm
decreases with temperature even at the lowest temperatures~\cite{Gio}. Also
the saturation value $R \sim 10\div 20$ k$\Omega$ (\ref{Rshort0}) predicted
here is consistent with that measured in \cite{Gio}.

Note that superconductivity in wires with radius $r_{0}\sim $5 nm is {\it not}
destroyed by finite size and level spacing effects; particles of radius down
to 1-3 nm turn superconducting~\cite{nano}. Finally, the $MS$ phase transition
discussed here is in many respects different from that in granular
wires~\cite{Gran}. In the latter case the role of disorder and the wire
thickness is irrelevant, and it is the onsite Coulomb interaction that drives
the transition into an insulating phase. For homogeneous wires, in contrast,
the transition is into a metallic state.

In conclusion, we have studied QPS's starting from microscopic theory and find
a measurable resistivity in superconducting ultrathin wires at temperatures
$T\ll T_{C}$, as well as a new superconductor to metal phase transition as a
function of the wire thickness.

We thank G. Blatter, R. Fazio, D. Geshkenbein, G. Sch\"{o}n, and
A. Tagliacozzo for encouragement. Support by the Humboldt Foundation, Deutsche
Forschungsgemeinschaft within SFB 195, the Swiss National Fonds, and NSF Grant
95-28535 is gratefully acknowledged.

\end{multicols}

\begin{references}
\bibitem{foot}  We will assume that the conductivity is equal to the normal
		state value inside the QPS core and vanishes outside at
		$T<\Delta_{0}$. This assumption can be violated due to
		non-equilibrium effects during the phase slip event and 
		the structure of the core -- depending on parameters -- could
		become more complicated~\cite{gozz}. This, however, does not
		lead to qualitative changes in our results.
\end{references}
\end{document}